\documentclass[a4paper,11pt]{article}

\usepackage{pos}
\usepackage{lineno}

\title{Refining the IceCube detector geometry using muon and LED calibration data}

\ShortTitle{Refining the IceCube detector geometry using muon and LED calibration data }

\author{The IceCube Collaboration \\{\normalsize \normalfont(a complete list of authors can be found at the end of the proceedings)}\\}

\emailAdd{matti.jansson@fysik.su.se}
\emailAdd{saskia.philippen@rwth-aachen.de}
\emailAdd{martin.rongen@fau.de}

\abstract{

The IceCube Neutrino Observatory deployed 5160 digital optical modules (DOMs) on 86 cables, called strings, in a cubic kilometer of deep glacial ice below the geographic South Pole. These record the Cherenkov light of passing charged particles. Knowledge of the DOM positions is vital for event reconstruction. While vertical positions have been calibrated, previous in-situ geometry calibration methods have been unable to measure horizontal deviations from the surface positions, largely due to degeneracies with ice model uncertainties. Thus the lateral position of the surface position of each hole is to date in almost all cases used as the lateral position of all DOMs on a given string. With the recent advances in ice modeling, two new in-situ measurements have now been undertaken. Using a large sample of muon tracks, the individual positions of all DOMs on a small number of strings around the center of the detector have been fitted.\\
Verifying the results against LED calibration data shows that the string-average corrections improve detector modeling. Directly fitting string-average geometry corrections for the full array using LED data agrees with the average corrections as derived from muons where available. Analyses are now ongoing to obtain per-DOM positions using both methods and in addition, methods are being developed to correct the recorded arrival times for the expected scattering delay, allowing for multilateration of the positions using nanosecond-precision propagation delays.

\vspace{4mm}
{\bfseries Corresponding authors:}
Matti Janson$^{1*}$, Saskia Philippen$^{2}$, Martin Rongen$^{3}$\\
{$^{1}$ \itshape Oskar Klein Centre and Dept. of Physics, Stockholm University, Sweden}\\
{$^{2}$ \itshape  III. Institute of Physics B, RWTH Aachen University, Germany}\\
{$^{3}$ \itshape Erlangen Centre for Astroparticle Physics, FAU Erlangen-Nürnberg, Germany}\\[4mm]
$^*$ Presenter

\ConferenceLogo{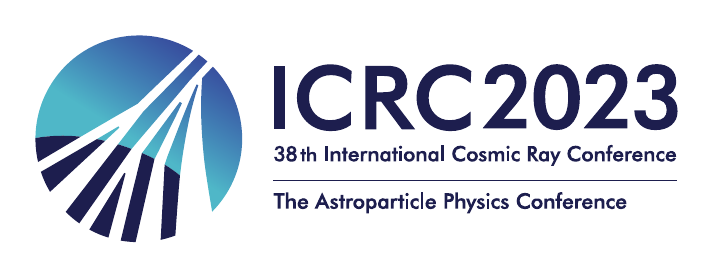}

\FullConference{The 38th International Cosmic Ray Conference (ICRC2023)\\ 26 July -- 3 August, 2023\\ Nagoya, Japan}
}

\begin{document}

\maketitle
%\begin{linenumbers}
\section{Introduction}\label{sec1}

The IceCube Neutrino Observatory is a neutrino detector instrumenting one cubic kilometer of deep, glacial ice at the geographic South Pole \cite{Aartsen:2016nxy}. It was built by drilling 86 holes of 60\,cm diameter each into the ice using a hot water drill \cite{EHWD}. Into each hole a cable holding 60 photosensors, called Digital Optical modules (DOMs), was lowered and left to freeze in place. 
Each DOM is capable of time-stamping the arrival time of individual photons, emitted as Cherenkov radiation from passing charged relativistic particles, to within 2\,ns uncertainty. The information available to event reconstructions in turn only consists of the amount, arrival time and DOM positions of the detected light.

To-date the detector geometry employed in simulation and reconstructions in almost all cases assumes all DOMs of a string to have the same lateral position as the center of the surveyed drill tower at the surface of the glacier. The depth of each string was initially measured by a pressure sensor located at a known distance below the last DOM and later updated to an accuracy of 0.2\,m by inter-string timing measurements using LED calibration data \cite{IceCube:2016zyt}.

Although gravity guided the drill to achieve near vertical holes, the unwinding of the drill hose from its spool induced some small lateral movements.
Since deployment, the overall detector has been shifting by 10\,m per year with respect to the underlying continent following the flow of the embedding ice. However, we assume the relative detector geometry to stay unperturbed by this ice movement. This assumption is justified over the time-scale of detector operation as the ice flow at the location of IceCube is believed to be dominated by basal sliding instead of plastic deformation, as inferred from inclinometer measurements\cite{IceCube:2016zyt} and acoustic sounding of a wet ice-rock interface~\cite{Peters2008}.

The orientation of the drill head was recorded throughout the drilling process. The integrated trajectories feature a maximum lateral deviation of 1.6\,m averaged over the 51 holes for which this data is still available. Comparing the calculated lateral positions for the downward and upward drill movement, the average  integrated error after a drill distances of 2500\,m was calculated to roughly 1\,m. Thus data from the drill head does not lend itself to improve on the detector geometry, but only sets an expectation that deviations from the surface position should rarely exceed 2 meters.

Since deployment, several attempts to calibrate the DOM lateral positions using trilateration of LED data~\cite{IceCube:2016zyt} as well as muon tomography have been undertaken. These were not incorporated into the default geometry as the studies were inconclusive, inconsistent or only applied to a small subset of strings. Geometry studies are generally challenging as only small timing or intensity differences are expected, and these are readily overwhelmed by systematic uncertainties associated with the ice optical modeling.

Following a number of recent improvements to the ice optical modeling \cite{tc-2022-174, TiltICRC23}, we present here two new attempts at measuring statistically significant shifts from the surveyed surface positions. The first method, described in section \ref{sec:muons}, employs a large set of muon tracks to find the most likely positions of individual DOMs on a subset of 18 strings within the more densely instrumented center of the detector. The second method, described in section \ref{sec:flasher}, fits string-average corrections for all strings of the detector to LED calibration data.
For the overlapping set of strings, these two methods for the first time achieve concordance as will be discussed in section \ref{sec:results}.

\section{Muon-based method}\label{sec:muons}

Using muons from cosmic rays to calibrate the geometry has already been proposed and tried with limited success in AMANDA\cite{dimageo}. While millions of muons give a high statistics sample, any muon-based method comes with the disadvantage of relying on a reconstruction and are therefore strongly affected by systematics.

With continuous improvements to the detector modeling and reconstructions, the prospects of muon tomography have recently improved. The overarching idea of the latest iteration is to find the DOM positions that yield the best goodness of fit in the description of the data with the muon reconstructions.
%the highest probability for the muon reconstructions to describe the data.

%In other words we use maximum likelihood \cite{10.1214/aoms/1177732360} to determine positions. 
%For the photon time PDF $p$, estimated using spline tables\cite{Whitehorn:2013nh}, track $track$, photon time $t$, DOM position ($x$, $y$, $z$) and noise level $n$ we get the following expression for the log likelihood.
%\begin{equation}
%    \log\Big(\prod_{\bar{t}}\prod_{t} \big( p(t |track, x, y, z) + n  \big) \Big)
%    \label{llhexpression}
%\end{equation}

%We can now search the space ($x$, $y$, $z$) for the maximum likelihood, yielding the best estimate of the DOM position. %For this analysis the noise level is set to $n=10^{-7}$ per ns. It is slightly lower than the measured rate but this analysis is is not very sensitive to the exact value.

\begin{figure}[h]
\centering
\begin{minipage}{.48\textwidth}
  \centering
  \includegraphics[width=\linewidth]{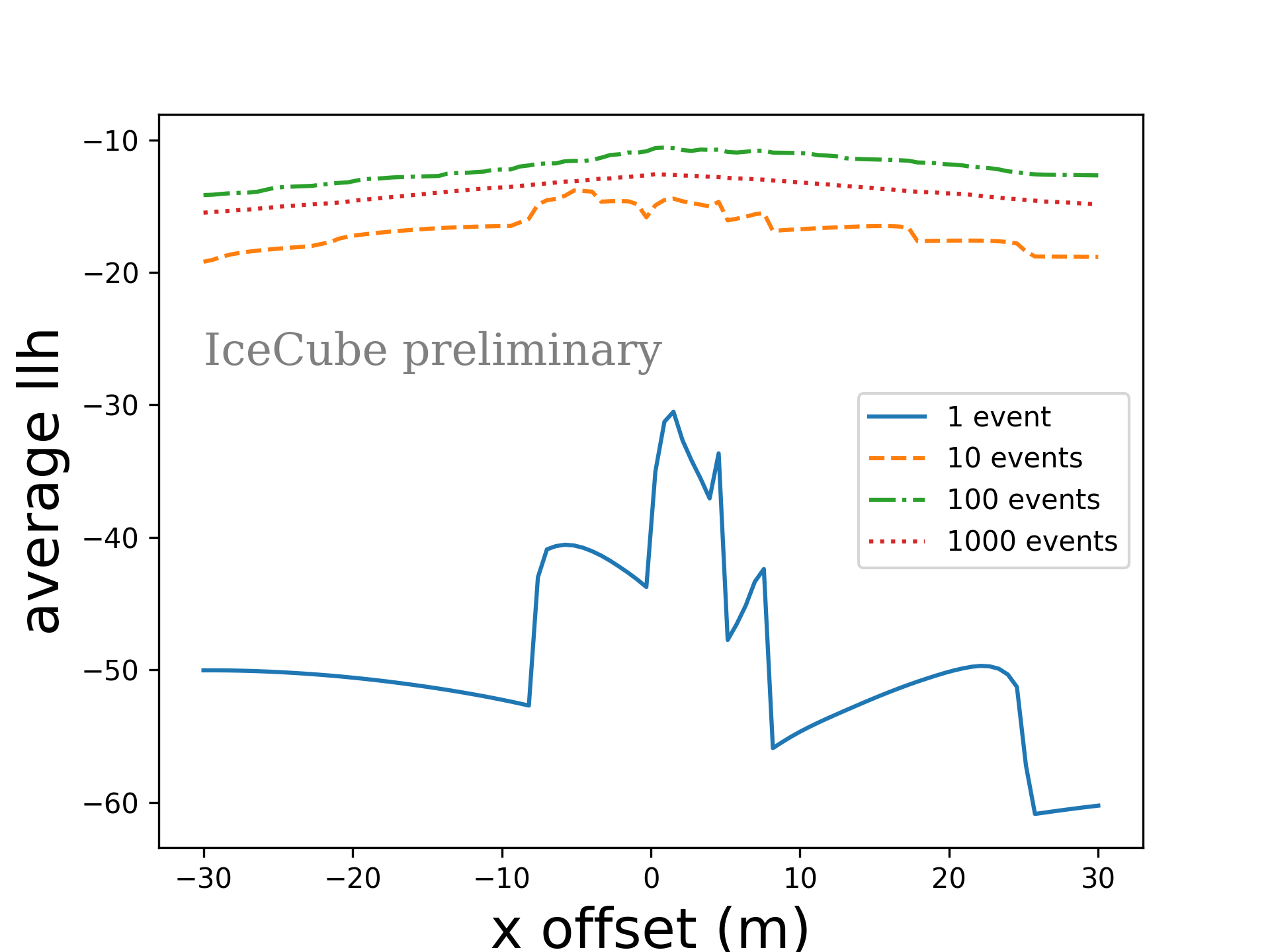}
  \captionof{figure}{Example progression of the likelihood space for one geometry coordinate of one DOM as more events are added. The average log likelihood over events is shown to make the different samples have comparable scale. }
  \label{fig:muonllh}
\end{minipage}
\hfill
\begin{minipage}{.48\textwidth}
  \centering
  \includegraphics[width=\linewidth]{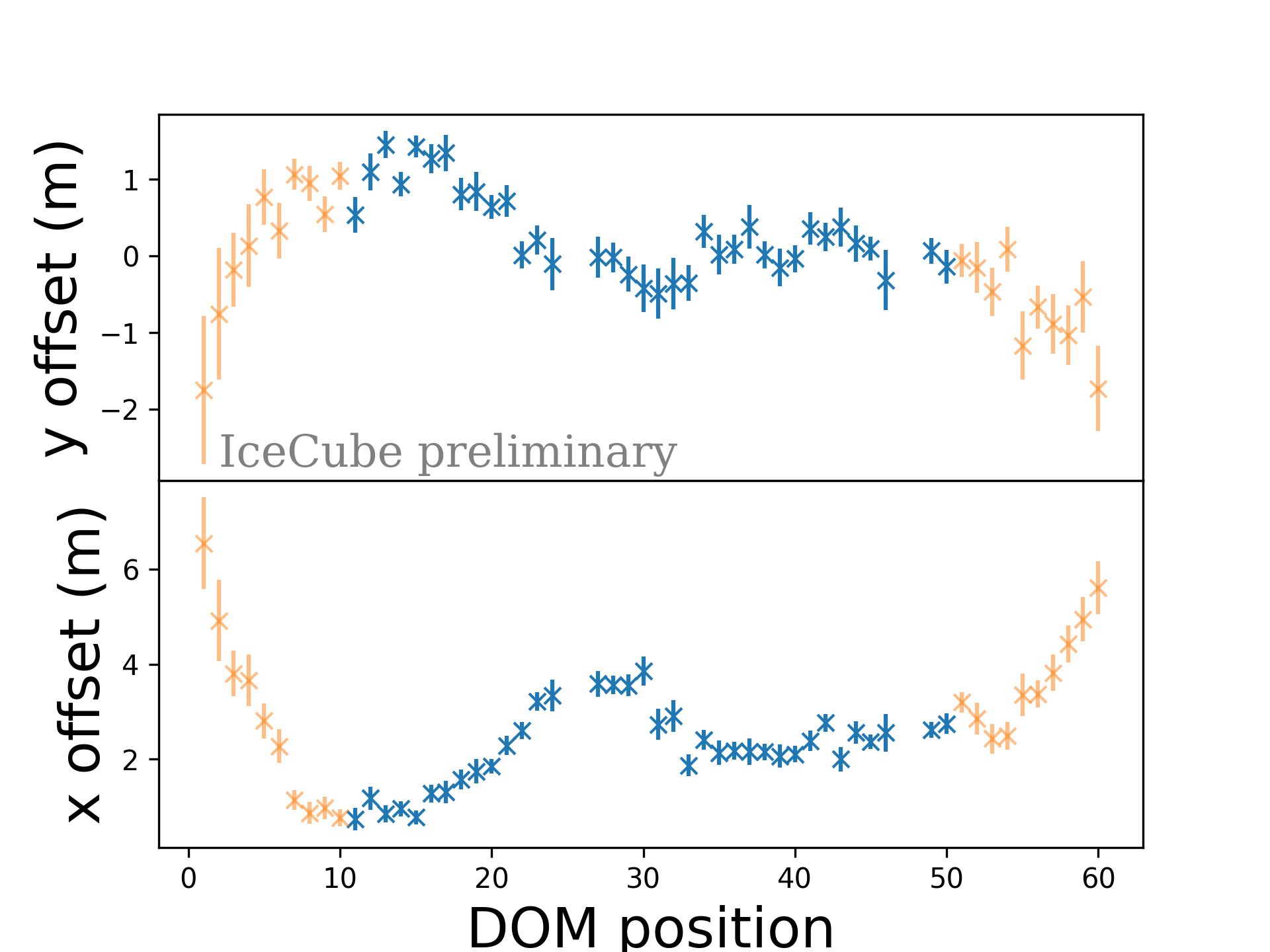}
  \captionof{figure}{Per-DOM lateral positions as fitted for string 44. Note the continuous development with depth. For the top and bottom DOMs large biases are systematically observed. Thus the 10 DOMs at the top and at the bottom are excluded when calculating the string-average position.}
  \label{fig:muonstring}
\end{minipage}
\end{figure}

Fitting all three coordinates for all DOMs simultaneously would require the optimization of over 15000 parameters which would be very computationally expensive. We may neglect the depth coordinate z from the fit as it is already known to within 0.2m. We may also restrict ourselves to fit the DOMs on one string at a time. 

%We expect DOMs on the same string to have similar errors, but as different tracks will use hits from different string during reconstruction and the deviation between strings are not correlated this systematic should cancel out to large degree. We continue by removing hits from the string in question before reconstructing the track. This makes the evaluation of likelihoods independent of hits on other DOMs on the same string, which would have been a problem as we expect any shifts for DOMs in the same string to be correlated.  It also allow us to split the job into 60 parallel jobs once the reconstruction has been completed. 

For this purpose the muon dataset is reconstructed without including data from the string in question and all other strings at their default positions. This assumes that the non-optimized default positions of the remaining strings do not on average bias muon reconstructions. 

Next the optimal position for each DOM on the string is fitted separately. For each DOM the probability of detecting the arrival times of the photons measured by this module is maximized as a function of the DOM's position assuming the previously reconstructed muon tracks. The expected photon arrival times are estimated using spline tables\cite{Whitehorn:2013nh}.  While these at the moment do not allow to encode the ice optical anisotropy~\cite{tc-2022-174}, nor the updated layer undulations~\cite{TiltICRC23}, they are otherwise based on the a recent ice model version.

Figure \ref{fig:muonllh} visualizes the evolution of the likelihood space for one coordinate of one example DOM. As only few photons are detected per event, the likelihood curves for a single event are rather discontinuous with many local maxima. As the event statistics increase the likelihood curve eventually becomes smooth, with a single clearly pronounced maximum.
Because of modeling errors carrying over from the initial reconstruction, Wilks's Theorem\cite{10.1214/aoms/1177732360} does not work for estimating error contours \cite{aef0ebfc-b11c-364d-9963-b4b4b9ecf97b} and instead bootstrapping~\cite{10.1214/aos/1176344552} over the muon sample is utilized. %In bootstrap the error of the estimate is estimated by resampling the data, allowing dublicates, creating new samples. And the distribution of the estimates using these samples are then used as the uncertainty of the original estimate. 

%\subsection{Z Estimatees and Edge Effects}
This method works well, without apparent systematic biases, in the center of the detector where symmetries cancel some systematic effects for horizontal shifts. For this reason we restrict ourselves to the 18 most central strings. When calculating the string-average position for comparison to the LED flasher method, the top and bottom 10 DOMs on each string are excluded for the same reason. %If we instead look at vertical shifts where there is a significant non symmetry as the PMT is pointed downwards we see large systematic effects. Similairly on the edges of the detector where all tracks appear on one side of the DOM see large systematic effects. 

\section{LED flasher-based method}\label{sec:flasher}

In addition to the photodetection hardware, each DOM is also equipped with 12 LEDs. These are arranged in pairs equally spaced around the equator of the pressure vessel, with one LED pointing horizontally outward into the ice and the the other LED pointing along anelevation angle of 48$^{\circ}$. The LEDs emit light with a wavelength of 405\,nm, with pulse durations configurable between 6\,ns and 70\,ns and reach intensities of up to $1.2 \cdot 10^{10}$ photons per pulse. During dedicated calibration runs, LEDs from a selected DOM are pulsed, and the arrival times of photons received in all other DOMs are recorded, in the process creating a light curve for each emitter-receiver pair of DOMs. This data is usually employed to measure the ice optical properties by iteratively simulating photon transport for different realizations of  assumed ice model parameters, and comparing the resulting light curves, with 25\,ns binning, to calibration data through a log-likelihood (LLH) minimization described in \cite{Flasher2013}.

\begin{figure}[h]
    \centering
    \includegraphics[width=\textwidth]{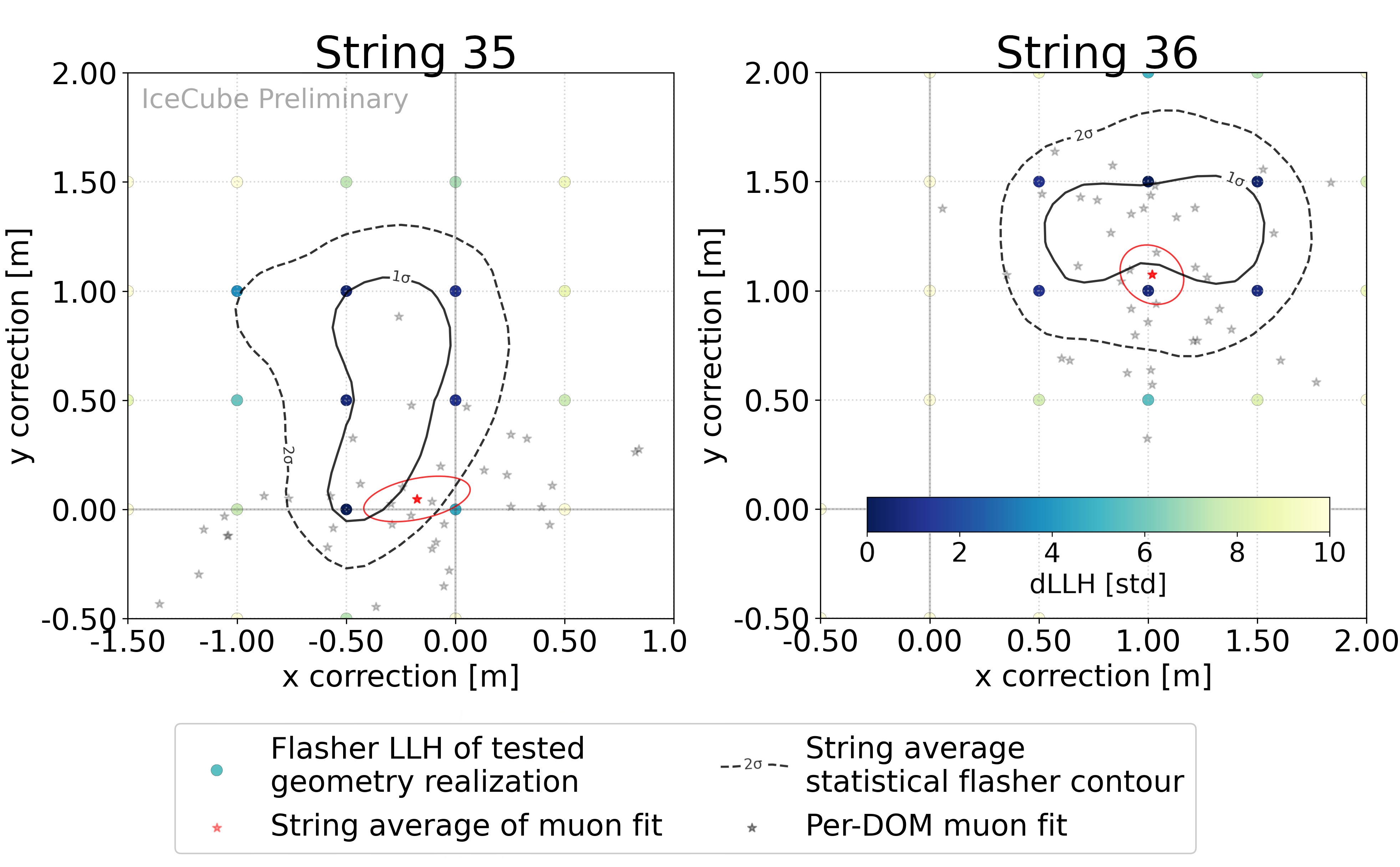}
    \caption{Visualization of the flasher-based geometry correction method and comparison of the string-average corrections for both methods, for two example strings. The grid of color dots shows the delta likelihood value of a flasher fit for a given simulated displacement of the emitter string, and the associated contours denote the statistical error. The red star and contour show the string-average correction of the muon analysis, obtained by averaging the per-DOM displacements (gray stars). The red contour denotes the statistical uncertainty on the mean.}
    %Comparison of fit performance between the muon and flasher analysis for individual example  strings. Colored circles show the delta likelihood value of a flasher fit for a given simulated displacement of the emitter string. The contours denote the statistical error only and are dominated by the simulation statistics. The gray stars denote the per-DOM fitted displacements for the same string resulting from the muon analysis. The red star and contour shows the string-average correction, obtained by averaging the per-DOM displacements. The red contour denotes the statistical uncertainty on the mean.
    \label{fig:llh}
\end{figure}

The same method is here employed to derive corrections to the default geometry, while keeping the ice optical properties fixed. As the ice model can easily be changed during photon transport simulation, we here employ a more recent ice model compared to the muon method, including the latest birefringence-based explanation of the ice optical anisotropy~\cite{tc-2022-174}. 

As was the case for the muon-based method, not all DOMs are varied and fitted simultaneously. Instead the fitting is performed one string at a time, with the other strings left at the default geometry and currently  only considering a single string-average correction for all DOMs on the string. In contrast to the muon method, where the string to be calibrated receives light, the string to be calibrated here acts as light emitter.

Figure \ref{fig:llh} shows likelihood landscapes for two example strings, which have also been considered in the muon-based method. Each circle represents one tested set of lateral corrections and is color coded according to the distance of the likelihood value from the best-fit realization. The employed likelihood~\cite{chirkin2013likelihood} accounts for the vastly smaller photon statistics in simulation compared to the experimental data. This induces fluctuations of the likelihood values compared to the expected paraboloid. The statistics-only uncertainty contours as shown account for this fluctuation by fitting a polynomial. As the likelihood does not conform to Wilks' Theorem the $\Delta LLH$ values for a given coverage have been calculated from the scatter observed by re-simulating one geometry realization several times. The contour sizes primarily reflect the employed simulation statistics, but are representative of the sensitivity of the analysis as a whole. 

In the provided examples both methods yield compatible results to within their claimed statistical uncertainty. The derived correction for string 35 is smaller than the hole diameter, while DOMs on string 36 appear to be on average offset by 1.8\,m. 

\section{Results}\label{sec:results}

\begin{figure}[h]
    \centering
    \includegraphics[width=0.9\textwidth]{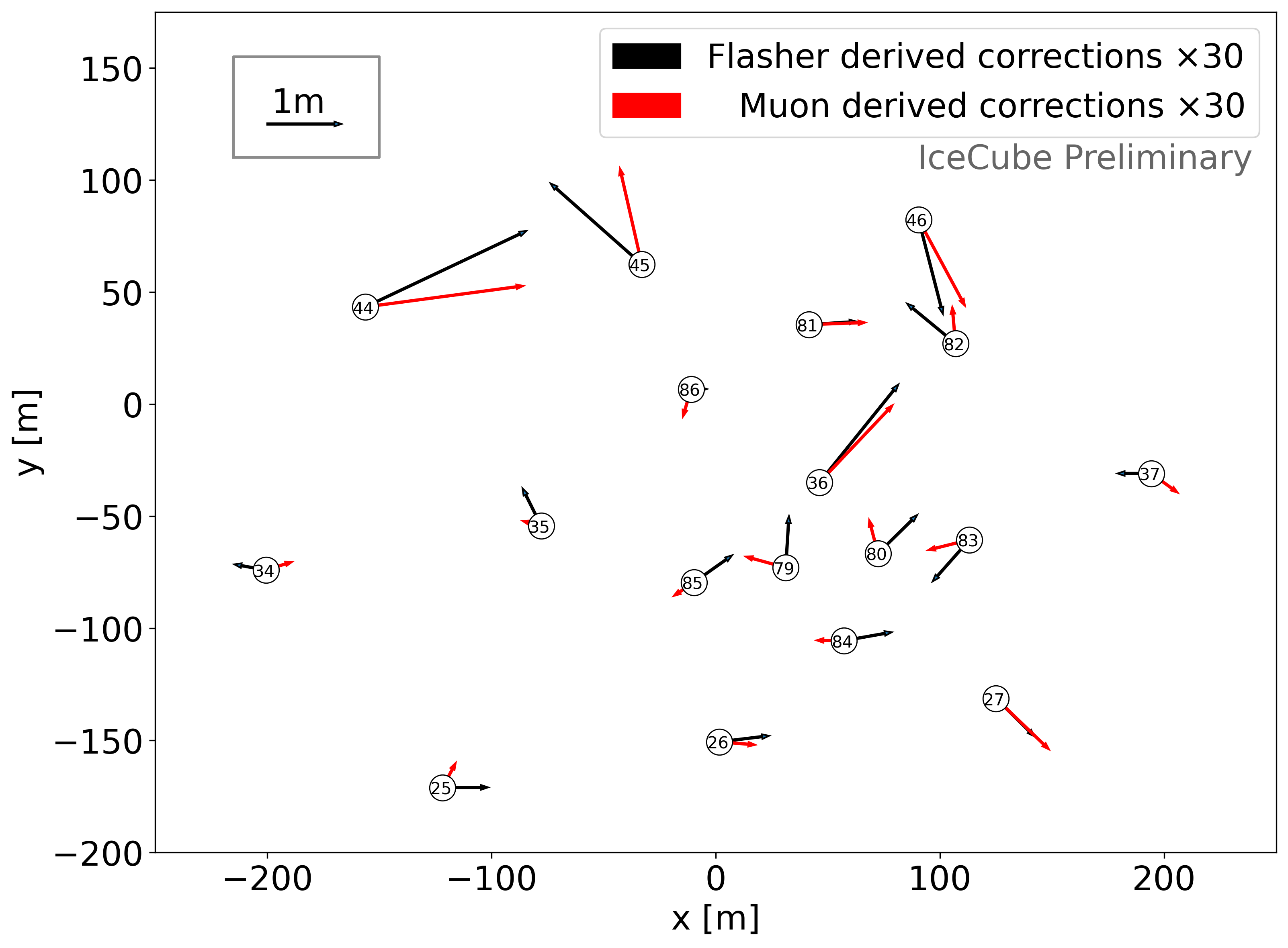}
    \caption{Geometry corrections in the DeepCore region of the IceCube footprint.  Strings in the default geometry are shown as circles with their number within. Vectors indicate the string-average geometry corrections as deduced by the two analyses and oversized by a factor 30 for better presentation.}       \label{fig:map}
\end{figure}

\begin{figure}[h]
    \centering
    \includegraphics[width=\textwidth]{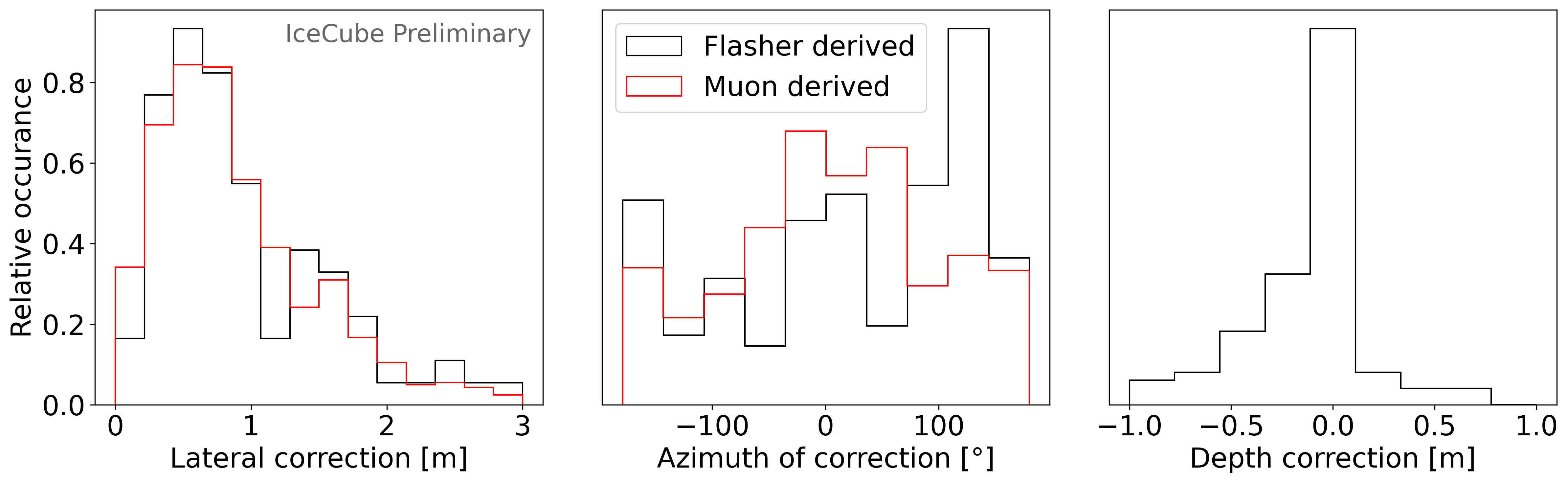}
    \caption{String-average corrections as derived by the flasher analysis compared to per-DOM corrections as derived by the muon analysis. See text for details.}
    \label{fig:hists}
\end{figure}

The map in Figure \ref{fig:map} provides a zoom into the central region of the detector, visualizing the corrections as derived by the two methods with the overlapping set of strings. Generally a good agreement between the methods is observed, in particular for the strings with the largest corrections. Averaged over all strings the LED flasher method yields string-average lateral corrections of 1.0\,m, with the distribution shown in Figure \ref{fig:hists}. For the overlapping strings, the mean distance between the fitted string position derived by the muon and LED flasher methods is 0.55\,m. Regardless of the individually derived statistical uncertainties, this may be considered the accuracy of the derived corrections and highlights that on-average significant corrections are found.

Depth corrections have only been derived using the flasher method. As seen in the right panel of Figure \ref{fig:hists} these scatter around the previously derived depth and with a standard deviation of 0.3\,m confirm the claimed accuracy of the previous flasher-derived depth corrections.

Without systematic biases to the geometry calibration one would assume the corrections to be randomly oriented. The central panel of figure \ref{fig:hists} shows a histogram of the azimuth directions of the lateral corrections weighted by the overall magnitudes of the corrections. While in particular the per-DOM corrections derived by the muon study strongly deviate from a uniform distribution, no correlations to known preferential directions such as the ice flow direction are clearly evident. 

\section{Conclusion and Outlook}\label{sec:outlook}

Two calibration methods deriving corrections to the currently assumed IceCube detector geometry, which for the lateral positions is simply based on the surface positions of the drill holes, have been presented. These methods are in good agreement, despite using independent data and different ice model assumptions. The derived string-averaged corrections are small ($\sim1$\,m), confirming the excellent performance by the employed hot water drill. Per-DOM corrections have so far been derived only using the muon method and remain under investigation.

While the concordance between methods is encouraging, further work is needed to assure the robustness of the preliminary results. This in particular entails round-trip tests, where the newly deduced geometry is used as a starting point and the fits are repeated. If, as assumed fitting individual strings without updating the positions of the surrounding strings yields unbiased results, one would expect on average no further corrections to be required. 
In addition systematics tests varying the ice optical modeling employed in flasher simulation,  introducing the new ice layer undulation maps and potentially enabling direct hole ice simulation~\cite{DimaICRC21}, may reveal yet unaccounted for biases inherent to both methods. Changing the underlying ice model assumptions in the muon method requires updated muon-spline-tables which are not readily available.

In parallel a new effort has been started to perform a purely timing-based trilateration analysis.  The LED flasher data are reprocessed with nanosecond binning to identify the least-scattered photon for each neighboring receiver DOM and applying further Monte-Carlo-derived corrections to account for the layered ice properties. 

%\end{linenumbers}
% Bibtex references:
\bibliographystyle{ICRC}
\bibliography{references}

\clearpage

%The following list of authors, affiliations and funding agencies will be updated at the day of submission. The following template is a placeholder generated via https://authorlist.icecube.wisc.edu/icecube on March 25, 2023 and will be updated.
\section*{Full Author List: IceCube Collaboration}

\scriptsize
\noindent
R. Abbasi$^{17}$,
M. Ackermann$^{63}$,
J. Adams$^{18}$,
S. K. Agarwalla$^{40,\: 64}$,
J. A. Aguilar$^{12}$,
M. Ahlers$^{22}$,
J.M. Alameddine$^{23}$,
N. M. Amin$^{44}$,
K. Andeen$^{42}$,
G. Anton$^{26}$
% ...
\\
\\
% ...
$^{12}$ Universit{\'e} Libre de Bruxelles, Science Faculty CP230, B-1050 Brussels, Belgium \\
% ...
$^{17}$ Department of Physics, Loyola University Chicago, Chicago, IL 60660, USA \\
$^{18}$ Dept. of Physics and Astronomy, University of Canterbury, Private Bag 4800, Christchurch, New Zealand \\
% ...
$^{22}$ Niels Bohr Institute, University of Copenhagen, DK-2100 Copenhagen, Denmark \\
$^{23}$ Dept. of Physics, TU Dortmund University, D-44221 Dortmund, Germany \\
% ...
$^{26}$ Erlangen Centre for Astroparticle Physics, Friedrich-Alexander-Universit{\"a}t Erlangen-N{\"u}rnberg, D-91058 Erlangen, Germany \\
% ...
$^{40}$ Dept. of Physics and Wisconsin IceCube Particle Astrophysics Center, University of Wisconsin{\textendash}Madison, Madison, WI 53706, USA \\
% ...
$^{42}$ Department of Physics, Marquette University, Milwaukee, WI, 53201, USA \\
% ...
$^{44}$ Bartol Research Institute and Dept. of Physics and Astronomy, University of Delaware, Newark, DE 19716, USA \\
% ...
$^{63}$ Deutsches Elektronen-Synchrotron DESY, Platanenallee 6, 15738 Zeuthen, Germany  \\
$^{64}$ Institute of Physics, Sachivalaya Marg, Sainik School Post, Bhubaneswar 751005, India \\
% ...

\subsection*{Acknowledgements}

\noindent
The authors gratefully acknowledge the support from the following agencies and institutions:
USA {\textendash} U.S. National Science Foundation-Office of Polar Programs,
U.S. National Science Foundation-Physics Division,
U.S. National Science Foundation-EPSCoR,
Wisconsin Alumni Research Foundation,
Center for High Throughput Computing (CHTC) at the University of Wisconsin{\textendash}Madison,
Open Science Grid (OSG),
Advanced Cyberinfrastructure Coordination Ecosystem: Services {\&} Support (ACCESS),
Frontera computing project at the Texas Advanced Computing Center,
U.S. Department of Energy-National Energy Research Scientific Computing Center,
Particle astrophysics research computing center at the University of Maryland,
Institute for Cyber-Enabled Research at Michigan State University,
and Astroparticle physics computational facility at Marquette University;
Belgium {\textendash} Funds for Scientific Research (FRS-FNRS and FWO),
FWO Odysseus and Big Science programmes,
and Belgian Federal Science Policy Office (Belspo);
Germany {\textendash} Bundesministerium f{\"u}r Bildung und Forschung (BMBF),
Deutsche Forschungsgemeinschaft (DFG),
Helmholtz Alliance for Astroparticle Physics (HAP),
Initiative and Networking Fund of the Helmholtz Association,
Deutsches Elektronen Synchrotron (DESY),
and High Performance Computing cluster of the RWTH Aachen;
Sweden {\textendash} Swedish Research Council,
Swedish Polar Research Secretariat,
Swedish National Infrastructure for Computing (SNIC),
and Knut and Alice Wallenberg Foundation;
European Union {\textendash} EGI Advanced Computing for research;
Australia {\textendash} Australian Research Council;
Canada {\textendash} Natural Sciences and Engineering Research Council of Canada,
Calcul Qu{\'e}bec, Compute Ontario, Canada Foundation for Innovation, WestGrid, and Compute Canada;
Denmark {\textendash} Villum Fonden, Carlsberg Foundation, and European Commission;
New Zealand {\textendash} Marsden Fund;
Japan {\textendash} Japan Society for Promotion of Science (JSPS)
and Institute for Global Prominent Research (IGPR) of Chiba University;
Korea {\textendash} National Research Foundation of Korea (NRF);
Switzerland {\textendash} Swiss National Science Foundation (SNSF);
United Kingdom {\textendash} Department of Physics, University of Oxford.
\end{document}